# Above-barrier heavy-ion fusion cross-sections using the relativistic mean-field approach: case of spherical colliding nuclei

M. V. Chushnyakova[a,*], M. Bhuyan[b,c], I. I. Gontchar[d], N. A. Khmyrova[d]

[a] *Physics Department, Omsk State Technical University, 644050 Omsk, Russia*
[b] *Department of Physics, Faculty of Science, University of Malaya, 50603 Kuala Lumpur, Malaysia*
[c] *Institute of Research and Development, Duy Tan University, 550000 Da Nang, Vietnam*
[d] *Physics and Chemistry Department, Omsk State Transport University, Omsk 644043, Russia*

**Abstract**

In the present work, the influence of the nuclear matter density on the DF potential and on the Coulomb barrier parameters is studied systematically for collisions of spherical nuclei. The value of the parameter $B_Z = Z_P Z_T / (A_P^{1/3} + A_T^{1/3})$ (estimating the Coulomb barrier height) varies in these calculations from 10 MeV up to 150 MeV. We have introduced self-consistent relativistic mean field (RMF) density in the present analysis. For the nucleon-nucleon effective interaction, the M3Y forces with the finite range exchange term and density dependence are employed. The above barrier fusion cross sections are calculated within the framework of the trajectory model with surface friction. Results are compared with the previous study in which the nuclear density came from the Skyrme Hartree-Fock (HF) calculations and with the high precision experimental data. This comparison demonstrates that i) agreement between the theoretical and experimental cross sections obtained with RMF and HF densities is of the same quality and ii) the values of the only adjustable parameter (friction strength) obtained with RMF and HF densities strongly correlate with each other.

\* Corresponding author.
E-mail address: maria.chushnyakova@gmail.com

## 1. Introduction

The fusion of two complex nuclei is an extremely interesting process: synthesis of elements with $Z = 105 \div 118$ is the most outstanding example. The best tool for describing this process theoretically is probably the time-dependent Hartree–Fock (TDHF) approach [1–3] within which the self-consistent evolution of the nucleon density and nucleus-nucleus interaction is obtained. However, this approach is extremely computing-time consuming and cannot account for fluctuations.

All other approaches are based on a nuclear part of the nucleus-nucleus interaction potential $U_n$ achieved within the framework of one or another model and subsequent simulation of the collision process. Among different approaches for obtaining $U_n$, averaging the effective nucleon-nucleon forces (*NN*-forces) with the frozen nucleon density distribution in the colliding nuclei seems to be both realistic and practical from the computing-time point of view. Such double-folding (DF) potentials use the M3Y nucleon-nucleon forces [4,5] or the Migdal forces [6]. The former approach is applied for calculating the potential $U_n$ in Refs. [7–11] whereas others prefer the Migdal *NN*-interaction [12–16]. Only recently the potentials $U_n$ resulting from these two kinds of *NN*-forces have been compared in detail for the case of spherical colliding nuclei [17,18].

The nucleon density distribution is the second important ingredient of the DF approach. Often for this density the Woods-Saxon ansatz is used [7,10–13,16,19]. However, the problem is that only the charge density distribution is experimentally measured [20,21] whereas for the DF approach the nucleon density is needed. It is more convenient to use the microscopically calculated nucleon density checking simultaneously whether the charge density resulting from the same calculations agree with the experimental data. The approach of this sort was applied in [22,23] for calculating the fusion (capture) excitation functions for reactions involving spherical nuclei. In [22] the nucleon density arrived from the Skyrme-Hartree-Fock calculations [24] with the SKP parametrization [25]. It turned out impossible to reproduce the high precision data [26] for the reactions with $^{16}$O. This failure was attributed to poor quality of the charge distribution obtained for $^{16}$O with the SKP parametrization. Therefore, the calculations had been repeated in [23] with the densities obtained on the basis of the Skyrme-Hartree-Fock calculations with the SKX parametrization [27]. In that case, both the charge density and fusion excitation functions were reproduced successfully.

It is known that the relativistic effects are ignored in the Skyrme-Hartree-Fock calculations. Therefore, in the present work, we try to see to what extent accounting for the relativistic effects can be of importance for the calculated fusion excitation functions. This is done by using the nuclear matter density arriving from the Relativistic Mean Field (RMF) approach with the NL3 parameter set [28]. For the sake of convenience, we refer through the paper to the results obtained with two different versions of nuclear densities as to the HF SKX and RMF NL3 (or SKX and NL3) results. In all calculations, we use the Paris M3Y *NN*-interaction of Ref. [5].

The paper is organized as follows. In Sec. 2 the RMF approach is shortly discussed. In Sec. 3 the dynamical model is described. In Sec. 4 the nuclear densities (Subsec. 4.1) and Coulomb barrier (Subsec. 4.2) are presented and the calculated cross sections are compared with the data (Subsec. 4.3). In Sec. 5 conclusions are formulated.

## 2. Relativistic mean-field approach for the nuclear densities

The fundamental theory of the strong interaction can provide a complete description of the nuclear equation of state entitled 'quantum chromodynamics' (QCD). At present, it is not conceivable to describe the complete picture of hadronic matter due to its non-perturbative nature. Hence, one needs to apply the perspective of an effective field theory at low energy, such as quantum hadrodynamics (QHD). The mean field treatment of QHD has been used widely to describe the properties of infinite nuclear matter and finite nuclei, where the nucleus is considered as a composite system of protons and neutrons interacting through the exchange of mesons and photons. The details of the relativistic mean field model and various parameter sets can be found in [29] and references therein. A typical relativistic Lagrangian density after several modifications of the original Walecka Lagrangian [30], to take care of various limitations for a nucleon-meson many body system, has the form [29–54]

$$\mathcal{L} = \bar{\psi}_i \{i\gamma^\mu \partial_\mu - M\}\psi_i + \frac{1}{2}\partial^\mu \sigma \partial_\mu \sigma - \frac{1}{2}m_\sigma^2 \sigma^2 -$$
$$-\frac{1}{3}g_2 \sigma^3 - \frac{1}{4}g_3 \sigma^4 - g_s \bar{\psi}_i \psi_i \sigma - \frac{1}{4}\Omega^{\mu\vartheta}\Omega_{\mu\vartheta} +$$
$$+ \frac{1}{2}m_\omega^2 V^\mu V_\mu - g_\omega \bar{\psi}_i \gamma^\mu \psi_i V_\mu - \frac{1}{4}\mathbf{B}^{\mu\vartheta} \cdot \mathbf{B}_{\mu\vartheta} -$$
$$- \frac{1}{2}m_\rho^2 \mathbf{R}^\mu \cdot \mathbf{R}_\mu - g_\rho \bar{\psi}_i \gamma^\mu \boldsymbol{\tau} \psi_i \cdot \mathbf{R}_\mu -$$
$$- \frac{1}{4}F^{\mu\vartheta}F_{\mu\vartheta} - e\bar{\psi}_i \gamma^\mu \frac{(1-\tau_{3i})}{2}\psi_i A_\mu \tag{1}$$

with vector field tensors

$$\begin{aligned} F^{\mu\vartheta} &= \partial_\mu A_\vartheta - \partial_\vartheta A_\mu, \\ \Omega^{\mu\vartheta} &= \partial_\mu \omega_\vartheta - \partial_\vartheta \omega_\mu, \\ \mathbf{B}^{\mu\vartheta} &= \partial_\mu \vec{\rho}_\vartheta - \partial_\vartheta \vec{\rho}_\mu. \end{aligned} \tag{2}$$

Here the fields for the $\sigma$-, $\omega$- and $\rho$- meson are denoted by $\sigma$, $\omega_\mu$, and $\rho_\mu$, respectively. The electromagnetic field is defined by $A_\mu$. The quantities $\Omega^{\mu\vartheta}$, $\mathbf{B}^{\mu\vartheta}$ and $F^{\mu\vartheta}$ are the field tensors for the $\omega^\mu$, $\rho^\mu$, and photon fields, respectively. From the above Lagrangian density, we obtain the field equations for the nucleons and mesons. These equations are solved by expanding the upper and lower components of the Dirac spinors and the boson fields in an axially deformed harmonic oscillator basis for an initial deformation $\beta_0$. The set of coupled equations is solved numerically by a self-consistent iteration method. The center-of-mass motion energy correction is estimated by the usual harmonic oscillator formula $E_{cm} = \frac{3}{4}(41\,A^{1/3})$. The quadrupole deformation parameter $\beta_2$ is evaluated from the resulting proton and neutron quadrupole moments, as

$$Q = Q_n + Q_p = \sqrt{\frac{16\pi}{5}}\frac{3}{4\pi}AR^2\beta_2. \tag{3}$$

The root mean square (rms) matter radius is defined as

$$\langle r_m^2 \rangle = \frac{1}{A}\int \rho(r_\perp, z) r^2 dr, \tag{4}$$

where $A$ is the mass number and $\rho(r_\perp, z)$ is the deformed density. The total binding energy and other observables are also obtained by using the standard relations, given in Ref. [50]. We apply the widely used NL3 [54] interaction parameter set for the present analysis. It is worth mentioning that the interaction parameters are able to describe the bulk properties of the nuclei reasonably good from the β-stable region to the drip-line [35,38,42,44–47]. To deal with the open-shell nuclei, one has to consider the pairing correlations in their ground as well as excited states [33,35,38,42,44–47]. The constant gap BCS approach is adopted for the present study and more details of the paring can be found in Ref. [31–40,42–54].

## 3. The trajectory fluctuation-dissipation model

*3.1. Dynamical equations and cross-sections*

The physical picture of our dynamical model is similar to that of Ref. [55] the detail description of the model can be found in [22]. The fictitious particle with the reduced mass runs experiencing the action of the conservative, dissipative, and random (fluctuating) forces. We study the collision process at the energies well exceeding the Coulomb barrier, therefore the quantum effects like tunneling and channels couplings are neglected. Here we consider the spherical nuclei which are rather stiff due to at least one (proton or neutron) closed shell. Therefore, we take into account only one degree of freedom corresponding to the radial motion. This motion is described by the dimensionless coordinate $q$ which is proportional to the distance between the centers of the projectile and target nuclei $R$. In [11] it was shown that accounting for the orbital degree of freedom could be ignored since it influenced the cross-sections within the framework of the statistical errors (typically 1%).

In [56] it was demonstrated that, in the collision process, the memory effects appear only near the contact configuration. In our modeling, this configuration is never reached, therefore we use the stochastic Langevin-type equation with the white noise and instant dissipation:

$$dp = (F_U + F_{cen} + F_D)dt + \sqrt{2D}\, dW, \tag{5}$$

$$dq = pdt/m_q, \tag{6}$$

$$F_U = -\frac{dU_{tot}}{dq}, \tag{7}$$

$$F_{cen} = \frac{\hbar^2 L^2}{m_q q^3}, \tag{8}$$

$$F_D = -\frac{p}{m_q} K_R \left[\frac{dU_n}{dq}\right]^2, \tag{9}$$

$$D = \theta K_R \left[\frac{dU_n}{dq}\right]^2. \tag{10}$$

Here $p$ denotes the linear momentum corresponding to the radial motion; $F_U$, $F_{cen}$, and $F_D$ are the conservative, centrifugal, and dissipative forces, respectively. The latter is related to the nucleus-nucleus strong interaction potential $U_n(q)$ via the surface friction expression [57,58]. $U_{tot}(q)$ is the total nucleus-nucleus interaction energy consisting of the Coulomb $U_C(q)$ and nuclear $U_n(q)$ parts; $L\hbar$ is the projection of the orbital angular momentum onto the axis perpendicular to the reaction plane; $m_q$ is the inertia parameter; $K_R$ denotes the dissipation strength coefficient; $D$ stands for the diffusion coefficient which is proportional to the temperature $\theta$. The random force is proportional to the increment of the Wiener process $dW$, the latter possesses zero average and variance equal to $dt$. Equations (5), (6) are solved numerically using the Runge-Kutta method (see details in [11,59]).

The capture cross-section is evaluated by means of the standard quantum mechanical formula (see e.g. [60])

$$\sigma_{th} = \frac{\pi\hbar^2}{2m_R E_{cm}} \sum_{L=0}^{L_{max}} (2L+1)T_L. \tag{11}$$

Here $m_R = m_n A_P A_T/(A_P + A_T)$, and $\hbar L_{max}$ is the maximal angular momentum above which the transmission coefficient becomes equal to zero.

*3.2. M3Y double-folding-potential*

A detailed description of the M3Y DF nucleus-nucleus potential can be found in many articles (see, e.g., [8,61,62]), therefore we give here only some basic formulas. The potential consists of the direct $U_{nD}$ and exchange $U_{nE}$ parts:

$$U_n(R, E_P) = U_{nD}(R, E_P) + U_{nE}(R, E_P). \tag{12}$$

The direct part reads:

$$U_{nD}(R, E_P) = g(E_P) \int d\vec{r}_P \int d\vec{r}_T \rho_{AP}(\vec{r}_P) F_v(\rho_{FA}) v_D(s) \rho_{AT}(\vec{r}_T). \tag{13}$$

Here $\vec{s} = \vec{R} + \vec{r}_T - \vec{r}_P$ corresponds to the distance between two points in the interacting nuclei, $v_D$ is the direct part of the effective NN-forces, the multiplier $g(E_P)$ in our case is very close to the unity ($E_P$ is the energy of the projectile per nucleon). The nucleon density $\rho_A(\vec{r})$ is the sum of the proton $\rho_p(\vec{r})$ and neutron $\rho_n(\vec{r})$ densities.

The function $F_v(\rho_{FA})$ for the density dependence of the NN-forces is taken from [61]:

$$F_v(\rho_{FA}) = C_v\{1 + \alpha_v \exp(-\beta_v \rho_{FA}) - \gamma_v \rho_{FA}\}. \tag{14}$$

The density at the middle-point between the centers of two nuclei is used for the argument of this function:

$$\rho_{FA} = \rho_{AP}(\vec{r}_P + \vec{s}/2) + \rho_{AT}(\vec{r}_T - \vec{s}/2). \tag{15}$$

The exchange part reads:

$$U_{nE}(R, E_P) = g(E_P) \int d\vec{r}_P \int d\vec{r}_T \rho_{AP}(\vec{r}_P; \vec{r}_P + \vec{s}) \times$$

$$\times F_v(\rho_{FA}) v_E(s) \rho_{AT}(\vec{r}_T; \vec{r}_T - \vec{s}) \exp(i\vec{k}_{rel}\vec{s}m_n/m_R). \qquad (16)$$

The direct $v_D$ and exchange $v_E$ parts of the effective *NN*-interaction consist of the Yukawa-type terms:

$$v_D(s) = \sum_{i=1}^{3} G_{Di}\left[\exp\left(-\frac{s}{r_{vi}}\right)\right]/\left(\frac{s}{r_{vi}}\right), \qquad (17)$$

$$v_E(s) = \sum_{i=1}^{3} G_{Ei}\left[\exp\left(-\frac{s}{r_{vi}}\right)\right]/\left(\frac{s}{r_{vi}}\right). \qquad (18)$$

We use here the values of $r_{vi}$, $G_{Di}$, and $G_{Ei}$ corresponding to [5] (the so-called Paris-forces): $G_{D1} = 11062$ MeV, $G_{D2} = -2537.5$ MeV, $G_{D3} = 0$, $G_{E1} = -1524.25$ MeV, $G_{E2} = -518.75$ MeV, $G_{E3} = -7.847$ MeV, $r_{v1} = 0.25$ fm, $r_{v2} = 0.40$ fm, $r_{v3} = 1.41$ fm. The coefficients for the *NN*-forces density dependence read $C_v = 0.3429$, $\alpha = 3.0232$, $\beta = 3.5512$ fm$^{-3}$, $\gamma = 0.5$ fm$^{-3}$ [61]. The influence of the version of the M3Y *NN*-forces (Paris or Reid) on the interaction barrier was studied in [63]: for the reaction $^{16}$O+$^{144}$Sm with the above coefficients for the density dependence, the difference is about 0.1 MeV. The impact of the density dependence version (i.e. the coefficients in Eq. (14)) is shown in [63] too: it is about 0.4 MeV (see Fig. 7 of that work).

**4. Results**

*4.1. Nuclear densities*

The nucleon densities for the nuclei involved in the present analysis resulting from the HF SKX and RMF NL3 approaches are displayed in Fig. 1. For the collision process, the tail of the nucleon density distribution is of primary importance, therefore the densities are shown in the logarithmic scale. Here and henceforth lines without symbols corresponds to the SKX density, boxes represent the results of calculations with the NL3 densities. One notices that the two approaches result in the densities significantly different at the tail for the heavier nuclei $^{144}$Sm and $^{204,208}$Pb whereas for the lighter nuclei the difference is much smaller.

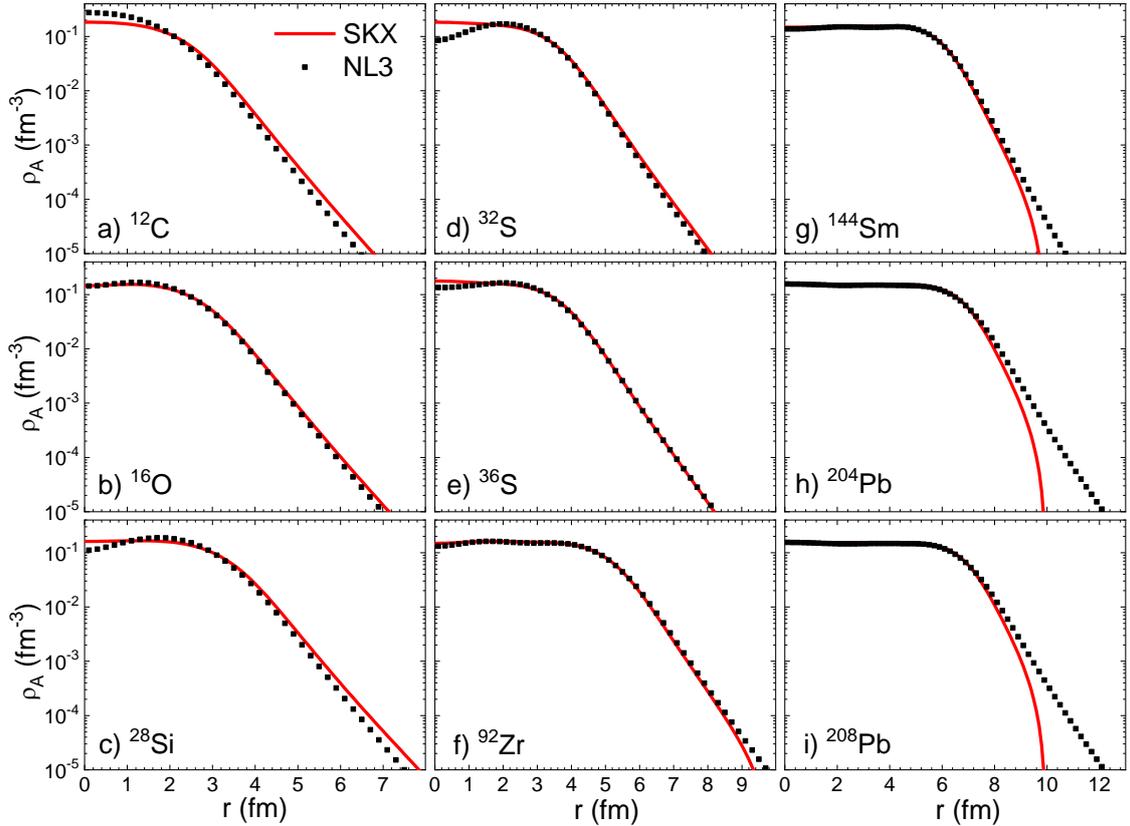

Fig. 1. The nucleon (matter) density versus the distance from the center of the nucleus for nine nuclei involved in the reactions considered in this work. Lines without symbols correspond to the HF SKX density, boxes represent the RMF NL3 densities.

*4.2. Coulomb barrier characteristics*

The impact of the different densities on the nucleus-nucleus potentials is illustrated qualitatively by Fig. 2. Here the potential is presented as the function of the center-of-mass distance for six reactions. One sees that for the lightest reactions (panels a and b) there is no apparent difference between the potentials. This is due to the comparatively small difference in the densities and small density overlap for the presented range of the center-of-mass distance as illustrated by Fig. 3 where the summed densities are shown for the barrier configuration. As the reaction becomes heavier, the difference between the SKX and NL3 potentials in Fig. 2 becomes more visible. The reason for that is stronger density overlap (see Fig. 3) and the stronger difference in the SKX and NL3 densities in the tail of the distribution for lead isotopes (see Fig. 1).

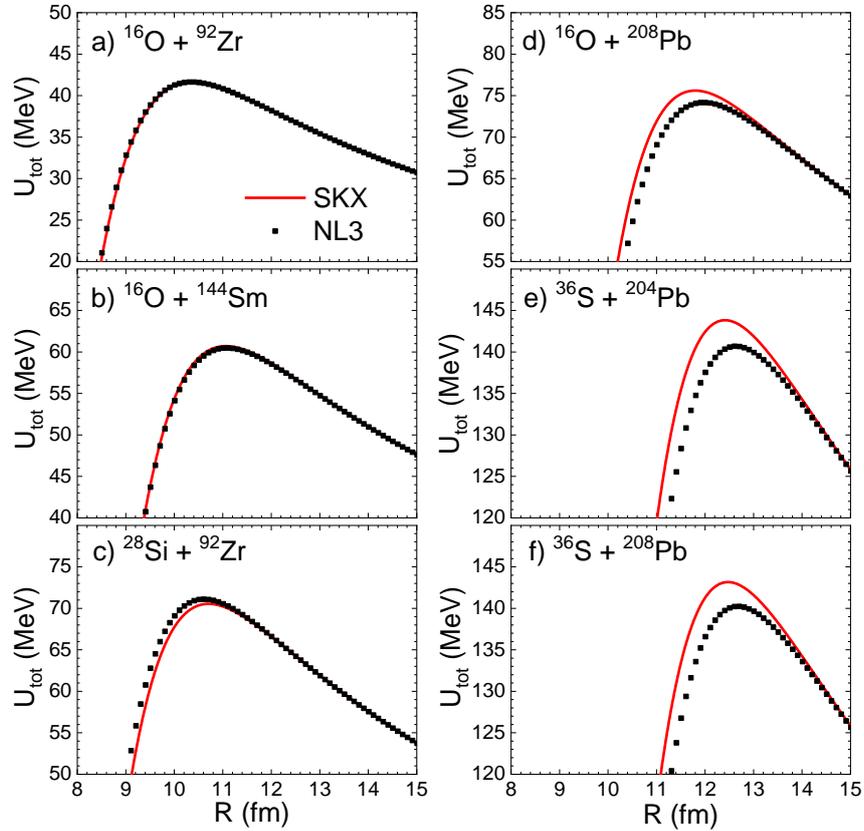

Fig. 2. The nucleus-nucleus interaction potentials versus the center-of-mass distance are shown for six reactions. Lines without symbols corresponds to the HF SKX density, boxes represent the results of calculations with RMF NL3 densities.

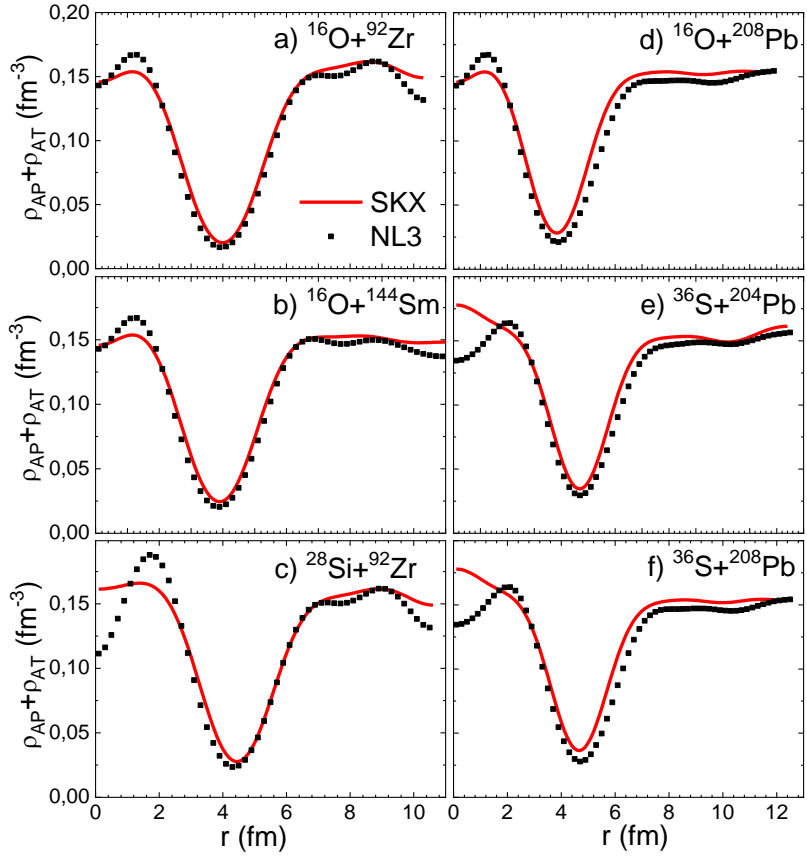

Fig. 3. The sum (projectile + target) matter densities for the barrier configuration. Notations are as in Fig. 1.

Let us now analyze parameters of the interaction barriers resulting from the DF approach with the two sorts of the matter density. A semi-quantitative impression on these parameters is provided by Fig. 4 where the barrier heights ($U_{B0}$, a) and radii ($R_{B0}$, b) at zero angular momentum are shown versus the approximate Coulomb barrier height $B_Z$:

$$B_Z = Z_P Z_T / \left(A_P^{1/3} + A_T^{1/3}\right) \text{ MeV}. \tag{19}$$

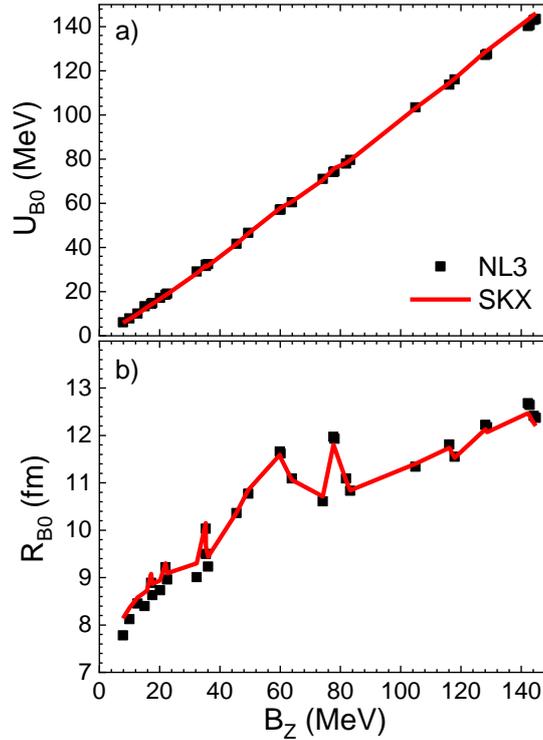

Fig. 4. The Coulomb barrier height (a) and radius (b) at the zero angular momentum versus $B_Z$ for all reactions listed in Table 1. Notations are as in Fig. 2.

One concludes from this figure that there is no striking difference in the barrier parameters when applying the non-relativistic and relativistic densities. Somewhat deeper insight into the dependencies of the barrier parameters upon the version of the density is provided by Figs. 5-8. In these figures the reduced barrier heights, $U_{B0}/B_Z$, and reduced barrier radii, $R_{B0}/R_{PT}$, are presented as the functions of $B_Z$ for $^{12}$C, $^{16}$O, $^{28}$Si, and $^{32,36}$S induced reactions, respectively. The approximate center-to-center distance at the contact point reads

$$R_{PT} = r_0\bigl(A_P^{1/3} + A_T^{1/3}\bigr), \qquad (20)$$

$r_0 = 1.2$ fm.

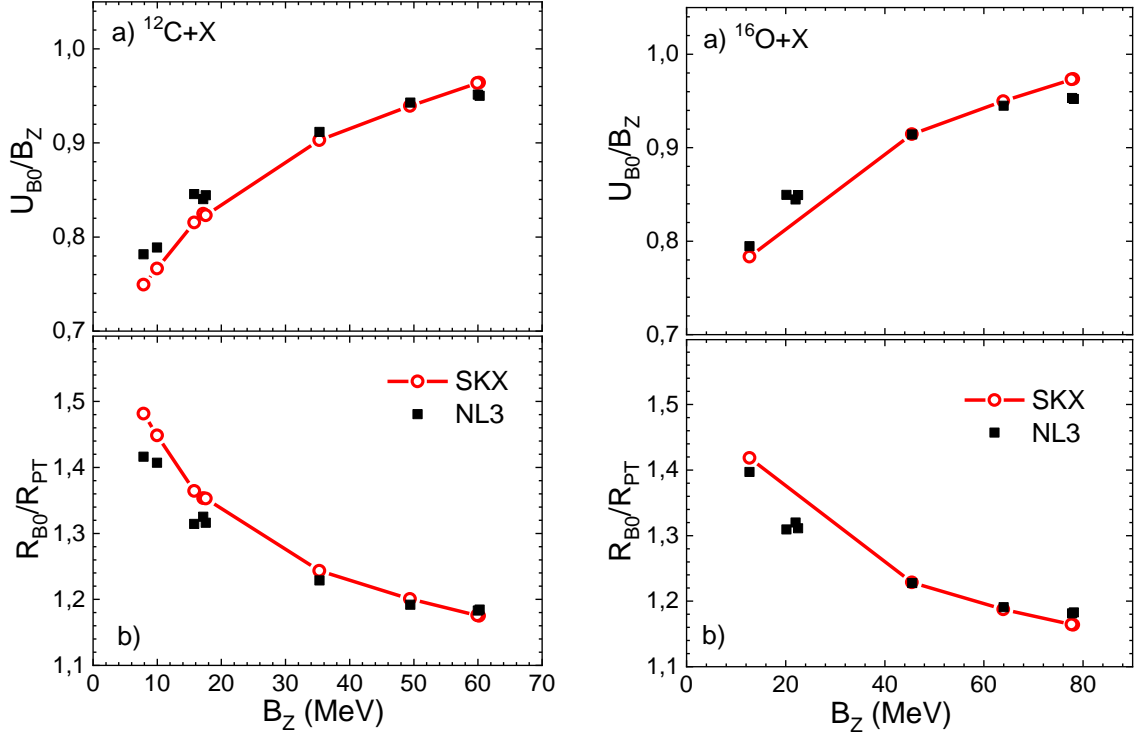

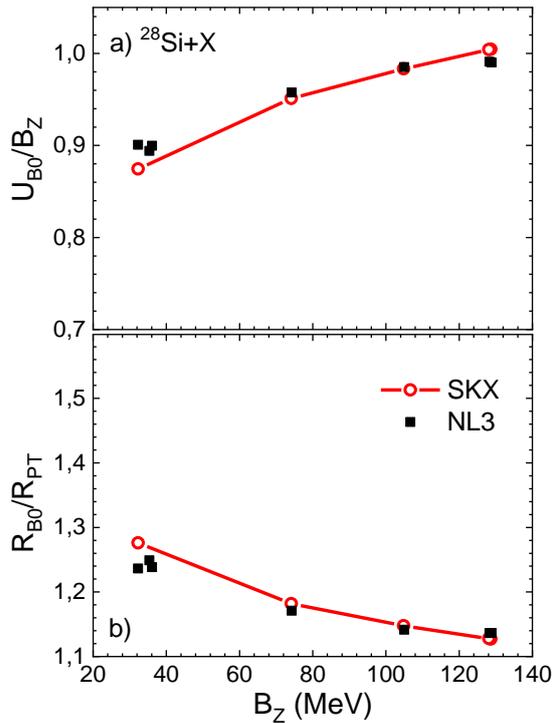
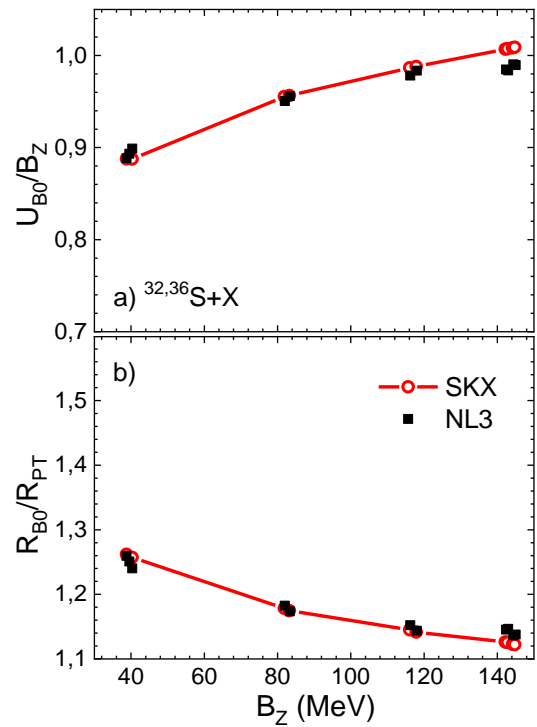

Fig. 5. The reduced Coulomb barrier height (a) and radius (b) versus $B_Z$ for $^{12}$C-induced reactions. Lines with symbols correspond to the SKX density, boxes represent the results of calculations with NL3 densities.
Fig. 6. Same as in Fig. 5 but for $^{16}$O-induced reactions.

Fig. 7. Same as in Figs. 5, 6 but for $^{28}$Si-induced reactions.   Fig. 8. Same as in Figs. 5-7 but for $^{36,32}$S-induced reactions.

Such presentation helps to get rid of the trivial increase in $U_{B0}$ and $R_{B0}$ with $B_Z$ due to geometry. In particular, we see that significant departure of $U_{B0}$ from $B_Z$ typical for lighter reactions disappears as the reaction becomes heavier (see panels an of Figs. 5-8). This at least partly results from the decrease of the reduced barrier radius in panels b) of that figures. These features are common for both types of density used. Let us now go over to the comparison of the barrier parameters resulting from the HF SKX and RMF NL3 approaches. In panels a) of Figs. 5-8, one sees that for the lighter reactions the NL3 barriers are several percents lower whereas for the heaviest reactions this tendency changes to the opposite. Naturally, for the barrier radii, the opposite is true. To provide the quantitative picture, we show in Tables 1-4 the fractional differences of the barrier heights and radii:

$$\xi_U = (U_{B0\ NL3} - U_{B0\ SKX})/U_{B0\ NL3}, \tag{21}$$

$$\xi_R = (R_{B0\ NL3} - R_{B0\ SKX})/R_{B0\ NL3}, \tag{22}$$

One sees that for each projectile at smaller $B_Z$ using the NL3 density results in the barrier which is several percents higher ($\xi_U > 0$) and more compact ($\xi_R < 0$) than the barrier obtained with the SKX density. As $B_Z$ increases for the given projectile nucleus, this tendency reverses although becomes weaker expressed. For example, for the reaction $^{16}$O+$^{28}$Si the value of $U_{B0\ NL3}$ is 2.3% larger and $R_{B0\ NL3}$ is 2.4% smaller whereas for the reaction $^{16}$O+$^{204}$Pb these figures are -2.1 and 1.6% respectively.

Table 1. Parameters of the $^{12}$C-induced reactions (the approximate Coulomb barrier height $B_Z$ and the sum of the projectile and target radii $R_{PT}$) and the fractional differences of the Coulomb barrier radii $\xi_R$ and heights $\xi_B$ defined by Eqs. (21), (22).

| Reaction | $B_Z$ (MeV) | $R_{PT}$ (fm) | $\xi_R$ (%) | $\xi_U$ (%) |
|---|---|---|---|---|
| $^{12}$C+$^{12}$C | 7.86 | 5.49 | -4,6 | 4,2 |
| $^{12}$C+$^{16}$O | 9.98 | 5.77 | -3.0 | 2,9 |
| $^{12}$C+$^{28}$Si | 15.77 | 6.39 | -3,8 | 3,7 |
| $^{12}$C+$^{36}$S | 17.17 | 6.71 | -2,1 | 1,9 |
| $^{12}$C+$^{32}$S | 17.57 | 6.56 | -2,8 | 2,6 |
| $^{12}$C+$^{92}$Zr | 35.27 | 8.16 | -1,2 | 1,1 |
| $^{12}$C+$^{144}$Sm | 49.40 | 9.04 | -0,74 | 0,51 |
| $^{12}$C+$^{208}$Pb | 59.89 | 9.86 | 0,60 | -1,2 |
| $^{12}$C+$^{204}$Pb | 60.17 | 9.81 | 0,77 | -1,3 |

Table 2. Same as in Table 1 but for the $^{16}$O-induced reactions.

| Reaction | $B_Z$ (MeV) | $R_{PT}$ (fm) | $\xi_R$ (%) | $\xi_U$ (%) |
|---|---|---|---|---|
| $^{16}$O+$^{16}$O | 12.70 | 6.05 | -1,5 | 1,5 |
| $^{16}$O+$^{28}$Si | 20.16 | 6.67 | -2,4 | 2,3 |
| $^{16}$O+$^{36}$S | 21.99 | 6.99 | -0,87 | 0,75 |
| $^{16}$O+$^{32}$S | 22.48 | 6.83 | -1,5 | 1,5 |
| $^{16}$O+$^{92}$Zr | 45.49 | 8.44 | -0,097 | 0,072 |
| $^{16}$O+$^{144}$Sm | 63.91 | 9.31 | 0,27 | -0,38 |
| $^{16}$O+$^{208}$Pb | 77.68 | 10.13 | 1,4 | -1,9 |
| $^{16}$O+$^{204}$Pb | 78.03 | 10.09 | 1,6 | -2,1 |

Table 3. Same as in Tables 1, 2 but for the $^{28}$Si-induced reactions.

| Reaction | $B_Z$ (MeV) | $R_{PT}$ (fm) | $\xi_R$ (%) | $\xi_U$ (%) |
|---|---|---|---|---|
| $^{28}$Si+$^{28}$Si | 32.27 | 7.29 | -3,2 | 3,0 |
| $^{28}$Si+$^{36}$S | 35.34 | 7.61 | -1,6 | 1,5 |
| $^{28}$Si+$^{32}$S | 36.06 | 7.45 | -2,3 | 2,2 |
| $^{28}$Si+$^{92}$Zr | 74.16 | 9.06 | -0,94 | 0,79 |
| $^{28}$Si+$^{144}$Sm | 104.9 | 9.93 | -0,53 | 0,35 |
| $^{28}$Si+$^{208}$Pb | 128.1 | 10.75 | 0,74 | -1,2 |
| $^{28}$Si+$^{204}$Pb | 128.7 | 10.71 | 0,82 | -1,3 |

Table 4. Same as in Tables 1-3 but for the $^{32,36}$S-induced reactions.

| Reaction | $B_Z$ (MeV) | $R_{PT}$ (fm) | $\xi_R$ (%) | $\xi_U$ (%) |
|---|---|---|---|---|
| $^{36}$S+ $^{32}$S | 39.53 | 7.77 | -0,72 | 0,74 |
| $^{36}$S+ $^{36}$S | 38.77 | 7.92 | -0,20 | 0,17 |
| $^{36}$S+ $^{92}$Zr | 81.88 | 9.38 | 0,36 | -0,37 |
| $^{36}$S+ $^{144}$Sm | 116.1 | 10.25 | 0,59 | -0,70 |
| $^{36}$S+ $^{208}$Pb | 142.2 | 11.07 | 1,7 | -2,1 |
| $^{36}$S+ $^{204}$Pb | 142.8 | 11.03 | 1,8 | -2,2 |
| $^{32}$S+ $^{32}$S | 40.32 | 7.62 | -1,4 | 1,4 |
| $^{32}$S+ $^{92}$Zr | 83.23 | 9.23 | -0,092 | 0,088 |
| $^{32}$S+ $^{144}$Sm | 117.9 | 10.10 | 0,17 | -0,30 |
| $^{32}$S+$^{208}$Pb | 144.2 | 10.92 | 1,2 | -1,7 |
| $^{32}$S+ $^{204}$Pb | 144.8 | 10.87 | 1,4 | -1,8 |

One can confront our results with respect to the system $^{16}$O+$^{208}$Pb with those obtained within the TDHF approach in Ref. [64]. In that work, the heights of the Coulomb barriers are shown calculated both self-consistently (accounting for the time evolution of the densities) $V_B^{DD}$ and within the frozen density approximation $V_B^{FD}$. Our calculations result in $U_{B0\;SKX} = 75.6$ MeV and $U_{B0\;NL3} = 74.2$ MeV. This is in reasonable agreement with $V_B^{FD} = 76.0$ MeV although the result achieved with the SKX-densities agrees better with the TDHF. It is interesting that the $V_B^{DD}$ varies in Fig. 11 of [64] from 75.0 up to 76.5 MeV in the collision energy interval considered in our work ($E_{cm} = 85.0 \div 109.5$ MeV). Thus, for this reaction, the barrier heights resulting from our fast calculations using both SKX and NL3 densities are in good agreement with the very computer time consuming self-consistent TDHF calculations.

*4.3. Capture cross-sections*

Let us now go over to the comparison of the evaluated capture (fusion) cross-sections with the experimental data. For this aim, we calculate the cross-sections $\sigma_{NL3}$, varying the value of the dissipation strength coefficient $K_R$ in Eqs. (9), (10) for a given reaction similar to Refs. [22,23]. At each value of $K_R$, the relative error

$$\chi_\sigma^2 = \frac{1}{v}\sum_{i=1}^{v}\left(\frac{\sigma_{iNL3} - \sigma_{iexp}}{\Delta\sigma_{iexp}}\right)^2 \quad (23)$$

is evaluated. Here $\sigma_{iNL3}$ is the value of the cross-section calculated at the particular value of $E_{cmi}$ using the RMF NL3 densities whereas $\sigma_{iexp}$ and $\Delta\sigma_{iexp}$ denote the experimental value of the cross-section and its error at the same collision energy. Searching for the minimum value of $\chi_\sigma^2$, we define the optimum value of the dissipation strength, $K_{Rm}$. In Table 5, the values of $K_{Rm}$ resulting from this study are compared with those obtained in [23]. The values of minimal $\chi_\sigma^2$ corresponding to these $K_{Rm}$ are shown in Table 5 as well. One sees, that the relative error of the NL3 calculations is typically smaller than of the SKX-calculations. Yet the optimal values of the dissipation strength are not significantly different in these two versions of the calculations.

The fusion excitation functions calculated with the optimal value $K_{Rm}$ are compared with the data in Fig. 9 where the ratio $r_\sigma$

$$r_\sigma = \sigma_{NL3}/\sigma_{exp}, \quad (24)$$

is plotted versus the ratio $U_{B0}/E_{cm}$ for 13 reactions listed in Table 5. Typical experimental errors of the data presented in this figure vary from 0.5% up to 5% whereas typical statistical error of the Langevin modeling is about 1%. Most of the reactions presented in Tables 1-4, are not included in the comparison with the experiment because the data are not accurate enough, uncertain (see [65]) or absent.

All 13 reactions are separated into four groups according to the projectile nucleus; each group is displayed in a special panel. One sees in Fig. 9 rather good agreement of the calculations with the data. Only for two points of 91 the ratio $r_\sigma$ is significantly beyond the 5%-interval around the unity (see panel a).

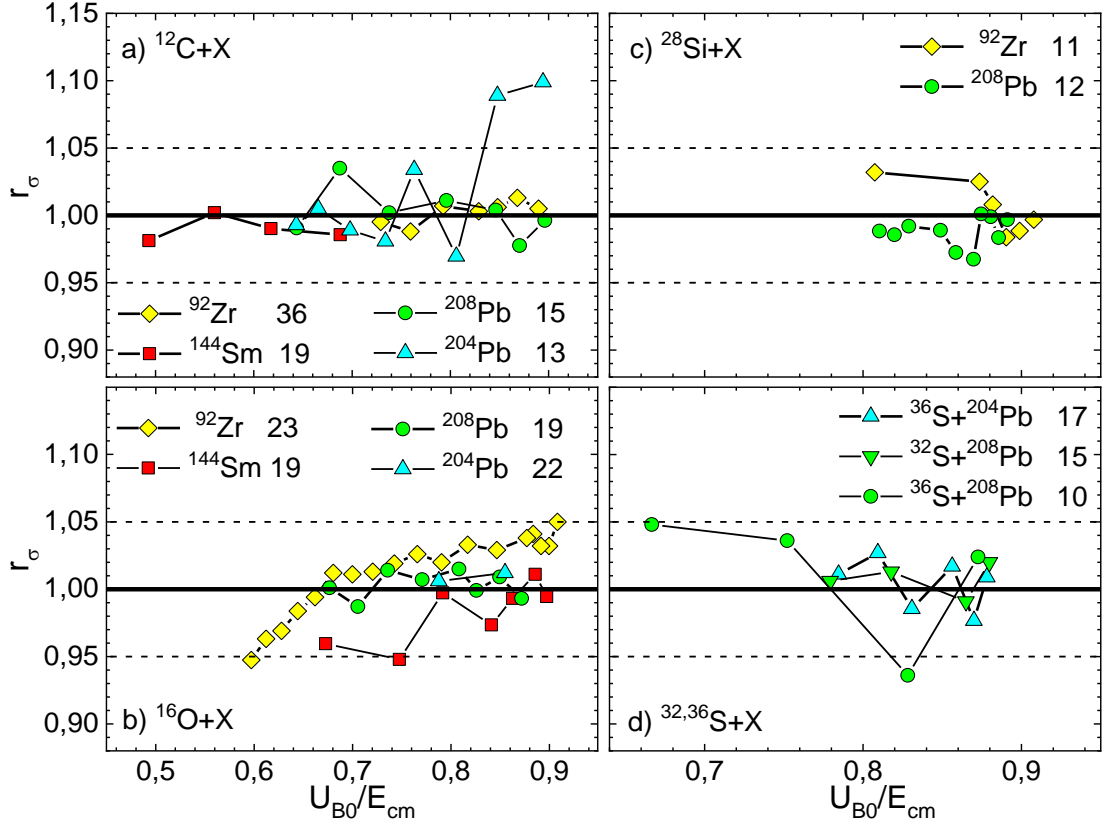

Fig. 9. The ratio $r_\sigma = \sigma_{NL3}/\sigma_{exp}$, as the function of $U_{B0}/E_{cm}$ for 13 reactions listed in Table 2. The references to the experimental data are given in Table 2; in many cases, the data are taken from [66].

Table 5. Parameters calculated using the RMF NL3 densities are confronted with those of Ref. [23] found using the HF SKX densities: the Coulomb barrier height $U_{B0}$ and radius $R_{B0}$, the optimum value of the dissipation strength $K_{Rm}$ providing the smallest relative error $\chi_\sigma^2$ for the cross-sections. In the last column, the reference to the high precision experimental capture (fusion) excitation function is shown.

| Reaction | $R_{B0\,NL3}$ (fm) | $R_{B0\,SKX}$ (fm) | $U_{B0\,NL3}$ (MeV) | $U_{B0\,SKX}$ (MeV) | $\chi_\sigma^2{}_{NL3}$ | $\chi_\sigma^2{}_{SKX}$ | $K_{Rm\,NL3}$ | $K_{Rm\,SKX}$ | Exp. data Refs. |
|---|---|---|---|---|---|---|---|---|---|
| $^{12}$C+$^{92}$Zr | 10.03 | 10.15 | 32.20 | 31.85 | 1.0 | 3.5 | 36 | 52 | [67] |
| $^{12}$C+$^{144}$Sm | 10.77 | 10.85 | 46.64 | 46.40 | 0.0 | 0.0 | 19 | 23 | [68] |
| $^{12}$C+$^{208}$Pb | 11.66 | 11.59 | 57.06 | 57.72 | 1.5 | 5.5 | 15 | 16 | [69] |
| $^{12}$C+$^{204}$Pb | 11.62 | 11.53 | 57.26 | 58.01 | 0.9 | 0.5 | 13 | 11 | [70] |
| $^{16}$O+$^{92}$Zr | 10.36 | 10.37 | 41.64 | 41.61 | 19 | 17 | 23 | 27 | [67] |
| $^{16}$O+$^{144}$Sm | 11.09 | 11.06 | 60.47 | 60.70 | 24 | 8.4 | 19 | 16 | [71] |
| $^{16}$O+$^{208}$Pb | 11.97 | 11.80 | 74.16 | 75.60 | 3.5 | 69 | 19 | 13 | [72] |
| $^{16}$O+$^{204}$Pb | 11.93 | 11.74 | 74.41 | 75.97 | 0.0 | 0.0 | 22 | 14 | [73] |
| $^{28}$Si+$^{92}$Zr | 10.61 | 10.71 | 71.10 | 70.54 | 4.6 | 2.8 | 11 | 19 | [67] |
| $^{28}$Si+$^{208}$Pb | 12.22 | 12.13 | 127.13 | 128.63 | 1.7 | 2.6 | 12 | 9 | [74] |
| $^{36}$S+$^{208}$Pb | 12.64 | 12.41 | 140.69 | 143.82 | 2.9 | 8.0 | 10 | 6 | [75] |
| $^{36}$S+$^{204}$Pb | 12.68 | 12.47 | 140.23 | 143.16 | 3.4 | 21 | 17 | 12 | [76] |
| $^{32}$S+$^{208}$Pb | 12.41 | 12.26 | 143.01 | 145.39 | 4.7 | 11 | 15 | 12 | [74] |

In Fig. 10 we show the values of the radial friction strength $K_{Rm}$ (symbols) that provides the minimal $\chi_\sigma^2$. It is seen that all symbols but one fall into the range between 10 and 25 zs·GeV$^{-1}$. On the other hand, the analytical approximation obtained Ref. [23] (thick line, SKX densities) does not significantly contradict to the present NL3 results.

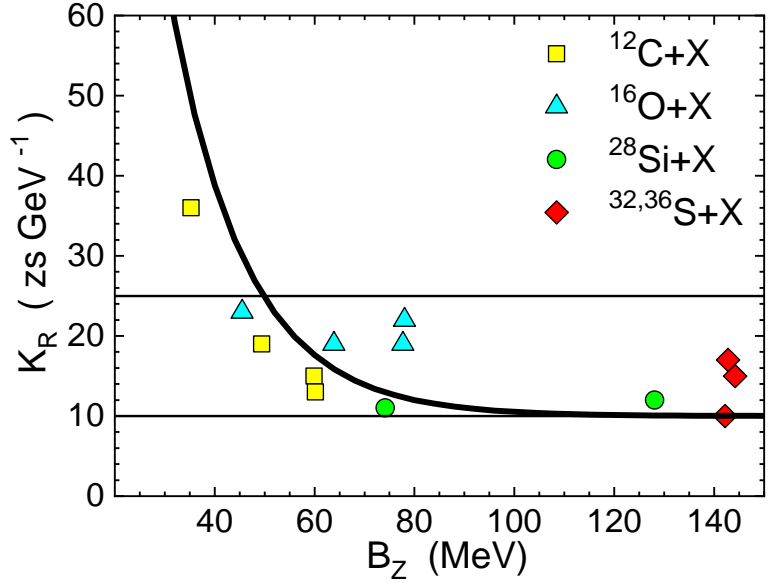

Fig. 10. The value of the radial friction strength $K_{Rm}$ that provides the minimal $\chi_\sigma^2$ versus $B_Z$ (symbols). The thick line represents $K_{Re}(B_Z)$ defined by Eq. (3) of Ref. [23] approximating results obtained with the SKX-densities. Horizontal lines indicate the range of $K_{Rm}$ between 10 and 25 zs·GeV$^{-1}$.

## 5. Conclusions

The double folding approach is one of the widely used methods for finding nucleus-nucleus interaction potential. Using this approach, in [23] it turned out possible to reproduce the experimental above barrier high precision fusion cross-sections with the nuclear densities obtained from the Skyrme-Hartree-Fock calculations with the SKX parametrization. For the effective nucleon-nucleon interaction the M3Y forces with the finite range exchange term and density dependence were employed in [23].

It is known that the relativistic effects were ignored in the Skyrme-Hartree-Fock calculations. Therefore, in the present work, we tried to see to what extent accounting for the relativistic effects can be of importance. This is done within the double folding scheme by adopting the nuclear matter densities from the relativistic mean-field approach with the NL3 parameter set [28]. The Paris M3Y $NN$-interaction was used. Calculations were performed for 35 reactions involving spherical nuclei, the value of the parameter $B_Z = Z_P Z_T / (A_P^{1/3} + A_T^{1/3})$ was varied from 10 MeV up to 150 MeV. Parameters of the Coulomb barrier (the height and the radius) obtained with the RMF densities turned out to be not significantly different from those obtained in [23] (see Tables 1-4).

Then, for 13 reactions, the above barrier fusion cross-sections were calculated within the framework of the trajectory model with surface friction [22]. Here one adjustable parameter, the dissipation strength coefficient $K_R$, was used. Comparison with the previous study and with the high precision experimental data demonstrated that i) agreement between the theoretical and experimental cross sections obtained with RMF and HF densities is of the same quality (see Table 5) and ii) the values of $K_{Rm}$ obtained with RMF and HF densities strongly correlate with each other (see Table 5).